\documentclass[10pt, aps, prd, preprintnumbers, superscriptaddress, nofootinbib, amsmath,amssymb]{revtex4-2}

\usepackage{style}
\usepackage{aas_macros}

\begin{document}

\preprint{DESY-24-118}

\title{The impact of cosmic variance on PTAs anisotropy searches}

\author{Thomas Konstandin\,\orcidlink{0000-0002-2492-7930}}
 \email{thomas.konstandin@desy.de}
\affiliation{Deutsches Elektronen-Synchrotron DESY, Notkestr. 85, 22607 Hamburg, Germany}

\author{Anna-Malin Lemke\,\orcidlink{0009-0005-3568-3336}}
 \email{anna-malin.lemke@desy.de}
\affiliation{II. Institute of Theoretical Physics, Universität Hamburg, Luruper Chaussee 149, 22761, Hamburg, Germany}

\author{Andrea Mitridate\,\orcidlink{0000-0003-2898-5844}}
 \email{andrea.mitridate@desy.de}
\affiliation{Deutsches Elektronen-Synchrotron DESY, Notkestr. 85, 22607 Hamburg, Germany}

\author{Enrico Perboni\,\orcidlink{0009-0002-5799-7625}}
 \email{enrico.perboni@desy.de}
\affiliation{Deutsches Elektronen-Synchrotron DESY, Notkestr. 85, 22607 Hamburg, Germany}

\begin{abstract}
Several Pulsar Timing Array (PTA) collaborations have recently found evidence for a Gravitational Wave Background (GWB) by measuring the perturbations that this background induces in the time-of-arrivals of pulsar signals. These perturbations are expected to be correlated across different pulsars and, for isotropic GWBs, the \emph{expected} values of these correlations (obtained by averaging over different GWB realizations) are a simple function of the pulsars' angular separations, known as the Hellings-Downs (HD) correlation function. On the other hand, anisotropic GWBs would induce deviations from these HD correlations in a way that can be used to search for anisotropic distributions of the GWB power. 
However, even for isotropic GWBs, interference between GW sources radiating at overlapping frequencies induces deviations from the HD correlation pattern, an effect known in the literature as ``\emph{cosmic variance}''.
In this work, we study the impact of cosmic variance on PTA anisotropy searches. We find that the fluctuations in cross-correlations related to cosmic variance can lead to the miss-classification of isotropic GWBs as anisotropic, leading to a false detection rate of $\sim 50\%$ for frequentist anisotropy searches. 
We also observe that cosmic variance complicates the reconstruction of the GWB sky map, making it more challenging to resolve bright GW hotspots, like the ones expected to be produced from a Supermassive Black Hole Binaries population. These findings highlight the need to refine anisotropy search techniques to improve our ability to reconstruct the GWB sky map and accurately assess the significance of any isotropy deviations we might find in it.
\end{abstract}

\maketitle
\tableofcontents
\clearpage
\section{Introduction}
Pulsar Timing Arrays (PTAs) monitor the arrival times of radiation pulses emitted by a collection of millisecond pulsars in our galactic neighborhood. In the pulsar's rest frame, these radiation pulses are emitted at very stable time intervals, allowing us to accurately predict the pulse's arrival times at the Earth location. A Gravitational Wave (GW) passing between the Earth and the pulsar will perturb these arrival times and induce deviations (also known as \emph{timing residuals}) from these predictions. One of the challenges of detecting these GW signals is disentangling them from other sources of time-of-arrival perturbations, such as intrinsic noise in the pulsar emission, propagation through the interstellar medium, and detector noise. This task is made possible because GWs affect the time-of-arrivals of all pulsars in a correlated and predictable fashion. Indeed, for a given pair of pulsars, the timing residuals induced by a passing GW are expected to be correlated, and the strength of these correlations is--\emph{on average}--a simple function of the pulsars' angular separation, known as the Hellings-Downs (HD) function~\cite{Hellings:1983fr}. It is important to stress that the HD correlation pattern is only obtained by averaging over pulsar pairs with a fixed angular separation, for the case of a single GW source, or averaging over many realizations of the Gravitational Wave Background (GWB), for a stochastic background of GWs. 

Recently, several PTA collaborations have found evidence for a stochastic signal with pulsar correlations consistent with the ones produced by a GWB~\cite{ng+23_gwb, EPTA:2023fyk, Reardon:2023gzh}. Future data sets will allow us to increase the significance of this finding and better characterize the properties of the GWB. For example, longer observation times and a more uniform distribution of pulsars in the sky will allow for better reconstruction of the sky distribution of the GWB power, potentially leading to the detection of anisotropy. Such a detection would provide insight into the origin of the GWB. In fact, the levels of anisotropy expected to be produced by cosmological sources (for a review of these sources see, for example, Ref.~\cite{NANOGrav:2023hvm}) are well below the reach of present and future PTAs~\cite{Caprini:2018mtu, LISACosmologyWorkingGroup:2022kbp, Cruz:2024svc}. In contrast, a population of Supermassive Black Hole Binaries (SMBHB) is expected to source a GWB with sizeable levels of anisotropy (see for example Refs.~\cite{Taylor:2013esa, Gardiner:2023zzr, Lemke:2024cdu}) that could be probed by future PTA data sets~\cite{Pol:2022sjn, Lemke:2024cdu, Depta:2024ykq}. Therefore, such a detection would provide strong evidence for an astrophysical origin of the observed background. 

Anisotropy searches exploit deviations from the HD correlations produced by anisotropic distributions of the GWB power to reconstruct and detect these anisotropies. However, even for isotropic GWBs, the interference between the GW radiation received from different sky directions (an effect that vanishes when averaged over many GWB realizations) can give rise to significant deviations from the HD correlations (see, for example, Refs.~\cite{Allen:2022dzg, Bernardo:2022xzl, Bernardo:2023bqx}). The total variance of these fluctuations can be broken down into two components: \emph{pulsar variance} and \emph{cosmic variance}. The first arises because, for a single realization of an isotropic GWB, pulsar pairs separated by the same angular position will have correlations that differ from the HD correlation depending on their position in the sky. This part of the total variance can be removed by binning the pulsar pairs by angular separation and taking the average of the correlations in the same angular-separation bin. Cosmic variance is the residual scatter that remains in the correlations after this averaging procedure, even for an infinite number of noise-free pulsars. 

In this work, we investigate the impact that the variance in the correlations of the pulsar signals, induced by these interference effects, have on anisotropy searches. Specifically, we study their impact on the frequentist search pipeline developed in Ref.~\cite{Pol:2022sjn} and adopted in the recent NANOGrav anisotropy search~\cite{NANOGrav:2023tcn}, and answer two main questions: First, can this correlation variance induce false anisotropy detections in current searches? Second, how do these fluctuations away from the mean affect our ability to reconstruct the GW sky and anisotropy detection prospects?

Before addressing these two questions, in Sec.~\ref{sec:timing_corr}, we review the formalism used to derive the pulsars' cross-correlations induced by a given realization of a GWB, derive the total and cosmic variance from simulations of an isotropic GWB, and break down all the effects contributing to the total correlation variance. In Sec.~\ref{sec:ani_search}, we review the current frequentist pipelines used to search for anisotropies in PTA data. Finally, in Sec.~\ref{sec:c_variance_impact}, we answer these two questions by quantifying the impact of the correlation variance on frequentist anisotropy searches. In Sec.~\ref{sec:conclusions}, we conclude. 

\section{Timing correlations in \texorpdfstring{PTA\MakeLowercase{s}}{PTAs}}\label{sec:timing_corr}
PTAs monitor the time of arrival (TOA) of radio pulses emitted by a population of galactic millisecond pulsars. A GW signal will appear in PTA data as a shift, $\delta t_a(t)$, in the arrival times of these radio pulses. Specifically, we can show that the shift induced by a generic metric perturbation in the TT gauge, $h_{ij}(t,\vec{x})$, can be written as (see, for example, Refs.~\cite{Maggiore:2018sht, Taylor:2021yjx}):
\begin{equation}\label{eq:z_gen}
    \delta t_a(t)=\frac{\hat p_a^i \hat p_a^j}{2}\int_{t-L_a}^t dt' h_{ij}\Big(t', (t-t')\hat p_a\Big)
\end{equation}
where $\hat p_a$ is the unit vector pointing from Earth to the $a$-th pulsar, and $L_a$ is the distance from Earth to the $a$-th pulsar in our array. The typical distance to the pulsars monitored in current PTA experiments is $L_a\sim$kpc. In this work, we used pulsar distances between 0.5 kpc and 1.5 kpc randomly generated drawing from a uniform distribution. These random distances are generated once and then used to derive all our results. We have checked that changing these distances within reasonable values does not significantly change our results. 

Assuming that the source(s) of the GW radiation are far away from the Earth-pulsar systems, we can decompose the GW metric perturbation in plane waves:
\begin{equation}\label{eq:gwb_metric}
    h_{ij}(t,\vec{x})=\sum_A\int_{-\infty}^{\infty}df\int_{S^2}d\hat\Omega\; \tilde h_A(f,\hat\Omega) e^{i 2\pi f(t-\hat\Omega\cdot\vec{x})}e_{ij}^A(\hat{\Omega})\,,
\end{equation}
where $f$ is the GW frequency, $\hat\Omega$ the direction of propagation of the plane waves, $A=+,\times$ labels the two GW polarizations, $e_{ij}^A$ are the GW polarization tensors, and $\tilde h_A(f,\hat\Omega)$ are two complex functions satisfying $\tilde h_A^*(f,\hat\Omega)=\tilde h_A(-f,\hat\Omega)$. By plugging this expansion into Eq.~\eqref{eq:z_gen}, we get\footnote{From here on, we drop the integration limits for the frequency and $\hat\Omega$ integral. Unless otherwise stated, the frequency integral is always over the entire real domain, while the $\hat\Omega$ integration is over a two-sphere.}
\begin{equation}\label{eq:z_plane}
    \delta t_a(t)=\int df\int d\hat\Omega\sum_A \tilde h_A(f,\hat\Omega) R^A_a(f,\hat\Omega) \dfrac{e^{2\pi i f t}}{2\pi i f}\,,
\end{equation}
where we have defined the response function $R_a^A(f,\hat\Omega)$ as:
\begin{equation}
    R^A_a(f,\hat\Omega)\equiv F^A_a(\hat\Omega) \left[1-e^{-2\pi i f L_a(1+\hat p_a\cdot\hat\Omega)}\right]\,,\qquad F^A_a(\hat\Omega)\equiv\frac{\hat p_a^i\hat p_a^j}{2(1+\hat\Omega\cdot \hat p_a)}e^A_{ij}(\hat\Omega)\,.
\end{equation}
The first term in the square brackets of the response function corresponds to the ``Earth term'', while the exponential in the square brackets gives the ``pulsar term''.

In a realistic PTA observation, on top of the GW signal, several noise sources will contribute to the observed TOA perturbations. However, the perturbations induced by a GW signal will feature specific correlations among pulsars that can be used to disentangle it from the noise. The correlation between the perturbations observed in a pair of pulsars, $\rho_{ab}$, can be defined as
\begin{equation}
    \rho_{ab}=\frac{1}{T}\int^{T/2}_{-T/2} dt\;\delta t_a(t)\delta t_b(t)\,,
\end{equation}
where $T$ is the total observation time. For simplicity, we will assume that all the pulsars in the array have the same total observation time. By substituting Eq.~\eqref{eq:z_plane} into the above expression, we find
\begin{equation}\label{eq:corr_broad}
    \rho_{ab}=\int \frac{df df'}{4\pi^2 ff'}\int d\hat\Omega d\hat\Omega'\sum_{AA'}\tilde h_A^*(f,\hat\Omega)\tilde h_{A'}(f',\hat\Omega')R_a^{A*}(f,\hat\Omega)R_b^{A'}(f',\hat\Omega')\,\sinc(\pi(f-f')T)\,.
\end{equation}
By writing the integrals in the previous equations as a sum over a set of discrete frequencies and equal-area pixels, we can show that $\rho_{ab}$ takes the form 
\begin{equation}\label{eq:corr_broad_pix}
    \rho_{ab}=\sum_{jj'}\sum_{kk'}\sum_{AA'}\left[\tilde h^{A*}_{kj}\tilde h^{A'}_{k'j'}\,R_{akj}^{A*}R_{bk'j'}^{A'}\,\sinc(\pi(f_j-f_{j'})T) - \tilde h^{A}_{kj}\tilde h^{A'}_{k'j'}\,R_{akj}^{A}R_{bk'j'}^{A'}\,\sinc(\pi(f_j+f_{j'})T)\right]\frac{\Delta\hat\Omega^2\Delta f^2}{4\pi^2f_jf_{j'}}+{\rm c.c.}\,,
\end{equation}
where $\Delta f$ is the frequency bin-size, $\Delta\hat\Omega=4\pi/N_{\rm pix}$ is the pixel area, and we have written the plane-wave coefficients as $\tilde h_A(f,\hat\Omega)\equiv h_A(f,\hat\Omega) e^{i\phi^A(f,\hat{\Omega})}$ (where $h_{A}(f,\hat\Omega)$ and $\phi_{A}(f,\hat\Omega)$ are real functions). The $k$-index labels the pixels of the equal-area tessellation of the sky, and the $j$-index labels the frequency bin centers, $f_j=(1+0.1j)/T$, where $j=0,1,\ldots,99$. With these conventions, we define $h^A_{jk}\equiv h^A(f_j,\hat\Omega_k)$, and $R_{akj}^A\equiv R_{a}^A(f_j,\hat\Omega_k)$, where $\hat\Omega_k$ is the unit vector pointing from the $k$-th pixel to the Earth location.
We obtain the sky tessellation by using the \texttt{HEALPix} package~\cite{healpix}, which controls map resolution via the $N_{\rm side}$ parameter, related to the map's pixel number, $N_{\rm pix}$, as $N_{\rm pix}=12 N_{\rm side}^2$. To derive the sky tessellation used to perform the $k$-sum in Eq.~\eqref{eq:corr_broad_pix}, we set $N_{\rm side}=16$, which corresponds to an angular resolution of $\sim 3.6^\circ$, well below the current resolution of $\sim 7.3^\circ$ of the NANOGrav 15-year data set~\cite{NANOGrav:2023tcn}. 

For any given realization of the GW, i.e. a given set of $\{h_{kj}^A,\,\phi_{kj}^A\}$, Eq.~\eqref{eq:corr_broad_pix} can be used to derive the correlations for any pair of pulsars uniquely. However, if we are interested in the signal from a stochastic GW background (GWB), we cannot make predictions about the specific values of $h_{kj}^A$ or $\phi_{kj}^A$. All we can do is make predictions about their statistical properties, which are set by the underlying mechanism producing the GWB. Since the GWB is expected to arise as a central-limit-theorem process, it is common to model the $\tilde h_{kj}^A$ coefficients as a multivariate Gaussian ensemble with zero mean and a two-point function given by 
\begin{equation}\label{eq:covariance}
    \langle \tilde h_A^*(f,\hat\Omega)\tilde h_{A'}(f',\hat\Omega')\rangle = \delta_{AA'}\delta(f-f')\delta(\hat{\Omega},\hat{\Omega}')H(\hat\Omega, f)\,,
\end{equation}
where $\delta_{AA'}$ arises from the assumption that the background is unpolarized,  $\delta(f-f')$ implies that the background is stationary in time, and $\delta(\hat{\Omega}, \hat{\Omega}')$ implies that the background is homogeneous. The GWB power spectrum, $H(\hat{\Omega},f)$, can be factorized as $H(\hat{\Omega},f)=H(f)P(\hat{\Omega},f)$, where the function $H(f)$ describes the spectral content of the GWB, and $P(\hat{\Omega},f)$ describes the distribution of the GWB power in the sky and is normalized such that $\int d\hat\Omega \, P(\hat\Omega,f)=4\pi$.
\begin{figure}[htbp]
    \centering
    \includegraphics[width=0.99\textwidth]{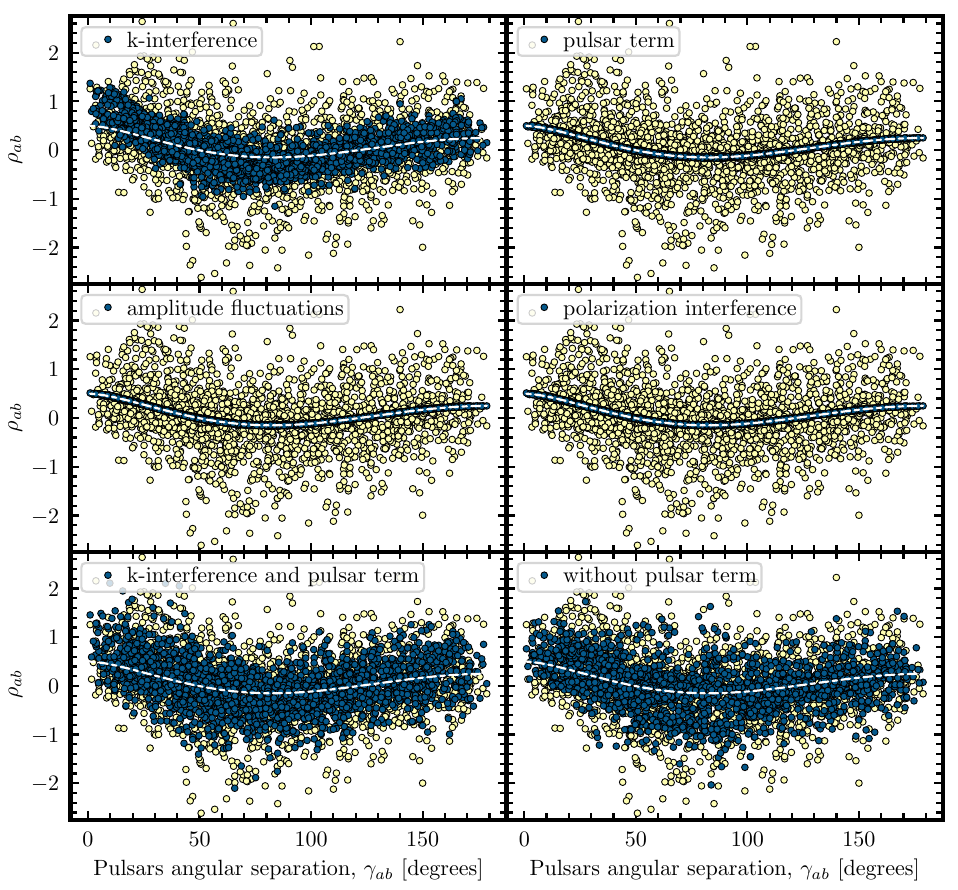}
    \caption{Pulsar cross-correlations induced by a single realization of an isotropic Gaussian GWB. In all the panels, the yellow dots represent the full correlations derived using Eq.~\eqref{eq:corr_broad_pix}, while the blue dots represent the cross-correlations  obtained including only
    some of the effects (mentioned in the bullet point list on the previous page) contributing to cross-correlation fluctuations away from the HD values. Specifically, in the \textbf{upper left panel}, the blue dots represent the cross-correlations that would be obtained only including the interference between the GW radiation received from different sky locations (i.e., preforming the replacement $\sum_{AA'}\to \sum_{A=A'}$, $h_{jk}^{A}\to\langle h_{jk}^{A}\rangle$, and $R_{akj}^{A}\to F_{ak}^{A}$ in Eq.~\eqref{eq:corr_broad_pix}); in the \textbf{upper right panel}, they represent the correlations obtained only including the effect of the pulsar term (which corresponds to the replacement $\sum_{AA'}\sum_{kk'}\to\sum_{A=A'}\sum_{k=k'}$ and $h_{jk}^{A}\to\langle h_{jk}^{A}\rangle$); in the \textbf{middle left panel}, they show the cross correlations obtained including only fluctuations in the GW amplitude of different pixels (which corresponds to the replacement $\sum_{AA'}\sum_{kk'}\to\sum_{A=A'}\sum_{k=k'}$ and $R_{akj}^{A}\to F_{ak}^{A}$); in the \textbf{middle right panel} they show the correlations obtained including only the contribution from polarization interference (which is equivalent to the replacement $\sum_{kk'}\to\sum_{k=k'}$, $R_{akj}^{A}\to F_{ak}^{A}$ and $h_{jk}^{A}\to\langle h_{jk}^{A}\rangle$); in the \textbf{bottom left panel} they show the correlations derived only keeping the contribution of the pulsar term and the interference of GW from different sky directions (which corresponds to the replacement $\sum_{AA'}\to \sum_{A=A'}$ and $h_{jk}^{A}\to\langle h_{jk}^{A}\rangle$); finally, in the \textbf{bottom right panel} we keep all but the pulsar term contribution (i.e., we perform the replacement $R_{akj}^{A}\to F_{ak}^{A}$). Notice that in all the panels we kept the effect of frequency interference. The white dashed lines in all panels show the HD correlation pattern.}
    \label{fig:scatter}
\end{figure}
The \emph{expected} cross-correlation coefficients, derived by using Eq.~\eqref{eq:covariance} to average over all possible realizations of the Gaussian ensemble, are given by:
\begin{equation}\label{eq:corr_avg}
    \qquad \qquad \qquad\langle\rho_{ab}\rangle= 2\sum_{j,k} H_j\left[F_{ak}^+F_{bk}^++F_{ak}^\times F_{bk}^\times\right]P_{kj}\,\frac{\Delta\hat\Omega\Delta f}{4\pi^2f_j^2}\qquad\qquad {\rm with}\; a\ne b
\end{equation}
where $H_j\equiv H(f_j)$, and $P_{k,j}\equiv P(\hat\Omega_k,f_j)$. In deriving this equation, we have used that, when $a\ne b$, the pulsar term is negligible for the diagonal terms of the $k$-sum so that we can approximate $R_{akj}^A\sim F_{ak}^A$. We can consider the expression in the square brackets of Eq.~\eqref{eq:corr_avg} as an \emph{average response function}, as it relates the GWB sky map, $P_{kj}$, to the \emph{average} (over the Gaussian ensemble) value of the cross-correlations. For an \emph{isotropic Gaussian background}, $P_{kj}=1\;\forall\;k,j$, and Eq.~\eqref{eq:corr_avg} reduces to well-known Hellings \& Downs (HD) correlation pattern. 
\begin{figure}[htbp]
    \centering
    \includegraphics{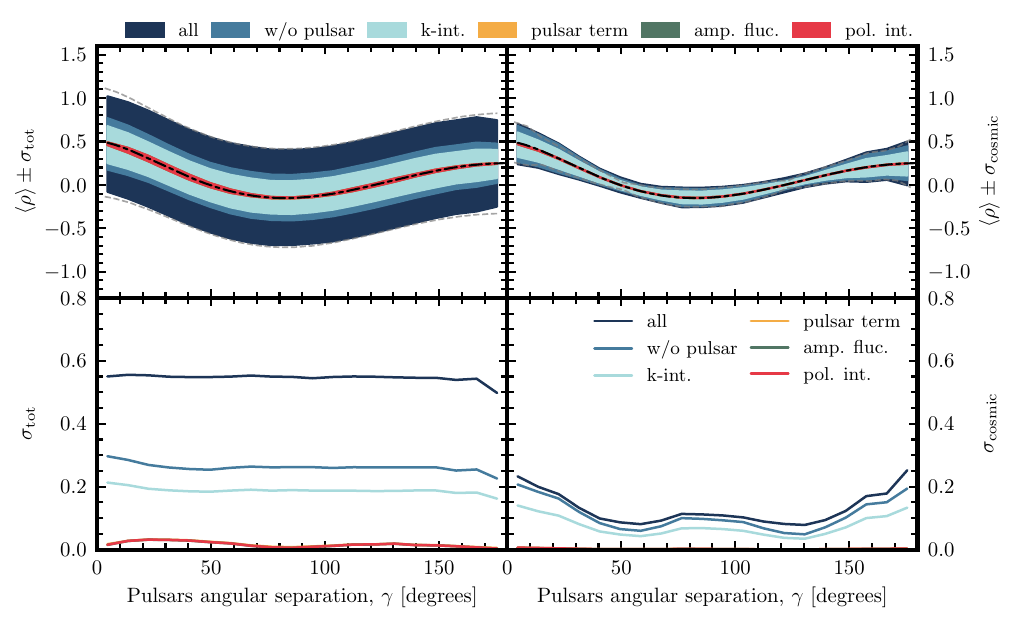}
    \caption{The upper left (right) panel shows the $\pm1\sigma_{\rm tot}$ ($\pm1\sigma_{\rm cosmic})$ regions around the mean cross correlation value. The lower left (right) panel shows the value of the total (cosmic) cross-correlation variance. The results are shown in 20 angular bins and were derived by averaging over 1000 realizations of an isotropic Gaussian GWB. 
    We derive these results by including all the contributions of fluctuations in cross-correlations (dark blue regions and lines), all but the pulsar term contribution (blue regions and lines), and only one of the contributions identified in the bullet points below Eq.~\eqref{eq:corr_avg} (light blue, yellow, green, and red regions and lines). 
    The gray dashed lines in the upper panels show the analytical results derived in Ref.~\cite{Allen:2022ksj}.
    Notice that in the upper panels the green and yellow bands are not visible because they completely overlap with the red band. Similarly, in the lower panels, the yellow and green lines are not visible because they almost perfectly overlap with the red line in the lower left panel, and with the x-axis in the lower-right panel. }
    \label{fig:variance}
\end{figure}
It is important to keep in mind that, in any given realization of the background (which is all we have access to), the cross-correlation values will fluctuate away from the mean given in Eq.~\eqref{eq:corr_avg}. Specifically, the cross-correlations induced by a single realization of an isotropic GWB will in general deviate from the HD values. Several effects contribute to these fluctuations away from the mean:
\begin{itemize}
    \item \emph{Source interference}: the GW signals received from different sky directions interfere with each other. This effect is captured by the off-diagonal elements of the sum over $k$ and $k'$ in Eq.~\eqref{eq:corr_broad_pix}. While this effect averages out when taking an ensemble average (since $\langle \tilde h_k^{A*} \tilde h_{k'}^{A'}\rangle\propto\delta_{kk'}$), in any single realization of the GWB this interference will contribute to the measured cross-correlation values.
    
    \item \emph{Pulsar term}: when $fL_a\gg 1$, the pulsar term gives a negligible contribution to the diagonal elements of the $k$-sum in Eq.~\eqref{eq:corr_broad_pix}. However, for the off-diagonal elements of this sum, the contribution of the pulsar term is non-negligible and increases the cross-correlations scatter around the mean by about a factor of $2$. This is consistent with the analytical findings of Ref.~\cite{Allen:2022dzg}.
    
    \item \emph{Pixel fluctuations}: any given realization of an isotropic ensemble will not appear as isotropic. Specifically, the values of the $h_k^A$ coefficients will not be a constant across the sky but fluctuate around zero with a variance set by Eq.~\eqref{eq:covariance}, and induce $\sim\mathcal{O}(1)$ fluctuations in the pixels amplitudes. These power fluctuations in the pixels add to the cross-correlation dispersion around the mean.
    
    \item \emph{Polarization interference}: the different polarizations of the GW signal interfere with each other. This effect is captured by the off-diagonal elements of the $A$, $A'$-sum in Eq.~\eqref{eq:corr_broad_pix}. While this effect averages out when taking an ensemble average (since $\langle \tilde h_{k}^{A*} \tilde h_{k’}^{A’}\rangle \propto \delta_{AA’}$), in any single realization of the GWB this interference will contribute to the measured cross-correlation values.

    \item \emph{Frequency interference}: the contributions from different GW frequencies interfere, introducing off-diagonal elements in the frequency sum over $f_{j}$ and $f_{j'}$ in Eq.~\eqref{eq:corr_broad_pix}. When averaging over an ensemble, this effect disappears (as $\langle \tilde h^{A*}_{kj} \tilde h^A_{kj'}\rangle \propto\delta_{jj'}$), but for individual realizations the interference terms, weighted according to $\sinc(\pi(f_j-f_{j'})T)$, are not negligible.
    
\end{itemize}
We show an example of these fluctuations away from the mean in Fig.~\ref{fig:scatter}, where we report the cross-correlations generated by a realization of an isotropic and Gaussian GWB. In this figure and the rest of this work, we normalize the cross-correlations by a factor
\begin{equation}\label{eq:norm}
    h^2 = \frac{4}{3\pi}\sum_jH_j\frac{\Delta f}{f_j^2}
\end{equation} 
such that the cross-correlations ensemble average matches the usual HD normalization (in this work, we follow the convention that the HD curve at zero separation is $1/2$).

We can quantify the magnitude of the cross-correlation fluctuations away from the mean by computing their variance~\cite{Allen:2022dzg, Allen:2022ksj}:
\begin{equation}\label{eq:var_tot}
    \sigma_{{\rm tot},ab}^2 = \langle \rho_{ab}^2\rangle-\langle\rho_{ab}\rangle^2\,,
\end{equation}
where $\langle\cdot\rangle$ indicates an average over the Gaussian ensemble given in Eq.~\eqref{eq:covariance}. Part of the total variance defined in Eq.~\eqref{eq:var_tot} arises from the fact that different pulsar pairs, separated by the same angle, respond differently to the same GWB realization. This contribution, commonly known as the \emph{pulsar variance}, can be removed by binning the pulsar pairs by separation angle, $\gamma$, and defining a bin-averaged correlation, $\Gamma(\gamma)$.\footnote{Specifically, the bin-averaged correlation function is defined as $\Gamma(\gamma) \equiv\sum_{ab}^{n_{\rm par}}
w_{ab} \rho_{ab}$, where the sum runs over all the $n_{\rm par}$ pulsar pairs with an angular separation that lies in the angular separation bin centered around $\gamma$, and $w_{ab}$ are dimensionless weights. The optimal choice of these weights, that takes into account the correlations between the $\rho_{ab}$, is described in Ref.~\cite{Allen:2022ksj}. In this work we use $w_{ab}=1/n_{\rm par}$.} The variance of this bin-averaged correlations, usually known as \emph{cosmic variance}, is given by 
\begin{equation}
    \sigma_{\rm cosmic}^2(\gamma)=\langle\Gamma(\gamma)^2\rangle-\langle\Gamma(\gamma)\rangle^2\,,
\end{equation}
and place a fundamental limit on our ability to predict the bin-averaged correlations measured by a PTA experiment. 
In Fig.~\ref{fig:variance}, we show the standard deviation of the cross-correlations coefficients across 1000 realizations of a statistically isotropic background, before (left panels) and after (right panels) binning the pulsars by angular separation, and averaging within each bin. From both Fig.\ref{fig:scatter} and Fig.~\ref{fig:variance}, we can see how the interference between GW radiation received from different sky locations and the pulsar term are the two main drivers of these deviations. On the other hand, the impact of pixel fluctuations and the interference between GW polarizations give negligible contributions to the variance of the correlations. As expected, our results for the total and cosmic variance of an isotropic Gaussian GWB agree with the ones derived analytically in Ref.~\cite{Allen:2022ksj}. One of the advantages of our numerical simulation is that it allows us to assess the contribution of each effect to the total variance.

\section{Frequentist anisotropy searches}\label{sec:ani_search}
The typical search for anisotropies consists of two main steps: first, the measured cross-correlations are used to reconstruct the map of the one GWB realization in our observable Universe; second, the statistical significance of any anisotropic distribution of power present in this map needs to be quantified. Practically, this second step consists of answering the following question: how likely is it for a realization of an isotropic GWB to produce the level of anisotropies observed in the reconstructed map?
Before discussing how the fluctuations in cross-correlations discussed in the previous section affect these two steps of anisotropy searches, in this section, we briefly review how frequentist searches for anisotropies are currently carried out.

\subsection{Sky reconstruction}\label{subsec:sky_rec}
To date, there have been only two searches for GWB anisotropies carried out by one of the regional PTAs. The first one was carried out by the European Pulsar Timing Array collaboration, using the Bayesian techniques developed in Ref.~\cite{Gair_2015}. The most recent one~\cite{NANOGrav:2023tcn}, carried out by the NANOGrav collaboration using its 15-year data set~\cite{NANOGrav:2023hde}, used two independent search strategies: one based on the Bayesian techniques discussed in Ref.~\cite{Taylor:2020zpk}, and another one based on a frequentist method developed in Ref.~\cite{Pol:2022sjn}. In this work, we will focus on the impact that cosmic and pulsar variance have on the latter search.

The frequentist search performed in Ref.~\cite{NANOGrav:2023tcn} assumes that the GW power has the same sky distribution across frequencies (i.e., $P(\hat\Omega,f)=P(\hat\Omega)$), and builds an estimator for this frequency-independent sky map, $\hat P(\hat\Omega)$, by maximizing the following likelihood function:
\begin{equation}\label{eq:likelihood}
    p(\bm{\rho}|\bm{P})=\frac{\exp[-\frac{1}{2}(\bm{\rho}-\bm{R}{\bm{P}})^T\mathbf{\Sigma}^{-1}(\bm{\rho}-\bm{R}{\bm{P}})]}{\sqrt{\det(2\pi\mathbf{\Sigma})}}\,,
\end{equation}
where $\bm{\rho}$ is a vector containing the cross-correlation for all the pulsar pairs, $\bm{P}$ is a vector containing the GWB power in each pixel, $\bm{\Sigma}$ is a diagonal matrix containing the cross-correlations uncertainties squared, and $\bm{R}$ is --up to normalization factors-- the average response function appearing in Eq.~\eqref{eq:corr_avg}:
\begin{equation}\label{eq:antenna_response}
    R_{k,ab}\equiv\frac{3}{2 N_{\rm pix}}\left[F_{a,k}^+F_{b,k}^++ F_{a,k}^\times F_{b,k}^\times\right]\,.
\end{equation}

Different parametrizations can be chosen for the sky map of the GWB power, $\bm{P}$. Here we discuss the most common ones, also adopted in Ref.~\cite{NANOGrav:2023tcn}:
\begin{itemize}
    \item \emph{Radiometer basis} -- Given the pixel representation used in previous equations, the most natural parametrization for the GWB power is one where each of the pixel values is taken as an independent parameter, $P_k$. In this basis, the pixel values that maximize the likelihood in Eq.~\eqref{eq:likelihood} can be found analytically as:
    \begin{equation}\label{eq:ML_solution}
        \hat{\bm{P}}=\bm{M}^{-1}\bm{X},
    \end{equation}
    where we have defined the Fisher information matrix, $\bm{M}=\bm{R}^{T}\bm{\Sigma}^{-1}\bm{R}$, and the ``dirty map", $\bm{X}=\bm{R}^{T}\bm{\Sigma}^{-1}\bm{\rho}$. Strictly speaking, the radiometer parametrization assumes that the GWB power is dominated by a single bright pixel. With this assumption, instead of reconstructing the GWB power over the entire sky, we can switch on one pixel at a time and reconstruct the power that each pixel would need in order to fit the measured cross-correlations. These optimal pixel values can be derived by replacing the Fisher matrix with its diagonal components in Eq.~\eqref{eq:ML_solution}. The angular resolution of the reconstructed maps is limited by the number of pulsar pairs, $N_{\rm pp}$, contained in the data set according to $N_{\rm pix}\lesssim N_{\rm pp}$~\cite{Romano:2016dpx}. For the case of the NANOGrav 15-year data sets, this limits the number of pixels to $N_{\rm pix}\lesssim 2211$. Because of this, and following the same convention of Ref.~\cite{NANOGrav:2023tcn}, we set $N_{\rm side}=8$ for the reconstructed maps.

    \item \emph{Spherical harmonic basis} -- In this basis, the GWB power is parametrized in terms of the coefficients, $c_{\ell m}$, of a linear decomposition in spherical harmonic functions:
    \begin{equation}
        P_k=\sum_{\ell=0}^{\ell_{\rm max}}\sum_{m=-\ell}^{m=\ell}c_{\ell m}Y_{\ell m}(\hat\Omega_k)\,.
    \end{equation}
    The resolution of the reconstructed maps is limited by the number of pulsars contained in the data set. In this case, we have $\ell_{\rm max}\lesssim \sqrt{N_{\rm psr}}$, where $N_{\rm psr}$ is the number of pulsars in the data set. For the NANOGrav 15-year data set this translates to $\ell_{\rm max}\sim8$, however, following the convention of Ref.~\cite{NANOGrav:2023tcn}, we set $\ell_{\rm max}=6$ to prevent over-fitting. 
    As for the radiometer basis, the maximum-likelihood values for the coefficients of this decomposition can be found analytically as
    \begin{equation}
        \hat{\bm{c}} = \bm{M}^{-1}\bm{X},
    \end{equation} 
    where the Fisher matrix and the dirty map appearing in this equation have been defined by using a modified version of the $\bm{R}$ matrix:
    \begin{equation}
        R_{(\ell m)(ab)} = \Delta\hat{\Omega}\sum_k \ Y_{lm}(\hat\Omega_k)\left[F_{ak}^+F_{bk}^++F_{ak}^\times F_{bk}^\times\right].
    \end{equation}
    \item \emph{Square-root spherical harmonic basis} -- The spherical harmonic parametrization allows the GWB power to assume negative values, which is --of course-- unphysical. A possible solution, proposed in Ref.~\cite{Taylor:2020zpk}, consists of parametrizing the square-root of the GWB using a spherical harmonic decomposition. This allows the square-root of the power to take negative values, while by construction always producing positive GWB power. Specifically, we write:
    \begin{equation}
        P_k = \left[P(\hat\Omega_k)^{1/2}\right]^2=\left[\sum_{L=0}^{L_{\rm max}}\sum_{M=-L}^{M=L}a_{LM}Y_{LM}(\hat\Omega_k)\right]^2\,,
    \end{equation}
    where, as for the linear spherical harmonic decomposition, we choose $L_{\rm max}=6$.
    The coefficients of the square-root decomposition, $a_{LM}$ can be related to the coefficients in the linear basis as
    \begin{equation}
        \displaystyle c_{\ell m} = \sum_{LM} \sum_{L^{\prime} M^{\prime}} a_{LM} a_{L^{\prime} M^{\prime}} \beta_{\ell m}^{LM, L^{\prime} M^{\prime}},\qquad\quad  \beta_{\ell m}^{LM, L^{\prime} M^{\prime}} = \sqrt{ \frac{(2L + 1) (2L^{\prime} + 1)}{4 \pi (2\ell + 1)}} C^{\ell m}_{LM, L^{\prime} M^{\prime}} C^{\ell 0}_{L0, L^{\prime} 0}\,,
    \end{equation}
    with $C^{\ell m}_{LM, L^{\prime} M^{\prime}}$ being Clebsch-Gordon coefficients. In this basis, the maximum-likelihood solution cannot be found analytically. Instead, we derived it by using numerical optimization techniques provided by the \texttt{LMFIT} package~\cite{matt_newville_2024_12785036}, as implemented in the \texttt{MAPS} package~\cite{Pol:2022sjn}.
\end{itemize}

In this work, we will only show results for maps reconstructed using the radiometer and square-root spherical harmonics parametrizations.

\subsection{Detection statistics and their calibration}\label{subsec:det_stat}
Once we have reconstructed a GWB sky map, the next step consists of defining a detection statistic to assess the level of compatibility between the reconstructed map and an isotropic GWB.
For maps reconstructed using the linear and square-root spherical harmonic parametrizations, a commonly adopted detection statistic --also used in the recent NANOGrav anisotropy search~\cite{NANOGrav:2023tcn}-- is the anisotropic signal-to-noise ratio (SNR), defined as the maximum likelihood ratio between the reconstructed GW sky and an isotropic sky:
\begin{equation}\label{eq:SNR}
    {\rm SNR} = \sqrt{2\ln\bigg[\frac{p(\bm{\rho}\vert\hat{\bm{P}})}{p(\bm{\rho}\vert \bm{P}_\text{iso})}\bigg]}\,,
\end{equation}
where $p$ is the likelihood function given in Eq.~\eqref{eq:likelihood}, $\hat{\bm{P}}$ is the recovered anisotropic sky map, and $\bm{P}_\text{iso}=1$ is a sky map with constant power across the sky. The statistical significance of a measured SNR is then assessed by comparing it with a null distribution obtained by simulating many realizations of the null hypothesis, i.e. many realizations of an isotropic GWB, and computing the associated SNR. Practically, in Ref.~\cite{NANOGrav:2023tcn}, this null distribution is constructed by generating, for each pulsar pair, mock correlations drawn from a normal distribution centered around the HD value and with a standard deviation given by the measured cross-correlation uncertainties. These mock correlations are then passed through the frequentist pipeline to reconstruct the GWB sky maps and the associated SNR. The collection of the SNRs obtained in this way constructs the null distribution, from which the $p$-value of the SNR measured on real data can be derived. 

For maps reconstructed using the radiometer basis, the measured power in each pixel, $\hat{P}_k$, can be used as a detection statistic. In this case, to establish if any of the reconstructed pixel values provide evidence for anisotropies, these values are compared with pixel-specific null distributions. Analogously to the SNR case, these null distributions are generated by calculating the detection statistic for many realizations of an isotropic GWB. Practically, Ref.~\cite{NANOGrav:2023tcn} generated these null distributions by drawing, for each pulsar pair, a random cross-correlation value from a normal distribution centered around the HD correlation and with a variance given by the measured cross-correlation uncertainties. The pixel values reconstructed for each of these realizations provide the pixel-specific null distributions.

\section{The impact of cosmic variance on anisotropy searches}\label{sec:c_variance_impact}
In this section, we discuss the impact that cross-correlations variance, discussed in Sec.~\ref{sec:timing_corr}, has on the anisotropy searches described in the previous section. Specifically, we will show how fluctuations in cross-correlations affect:

\begin{itemize}
    \item \emph{Detection statistics calibration}: as discussed in the previous section, when deriving null distributions for the detection statistics, current analyses assume that --up to reconstruction uncertainties-- the cross-correlations produced by an isotropic GWB are given by the HD correlations. However, as discussed in Sec.~\ref{sec:timing_corr}, any given realization of an isotropic GWB will produce cross-correlations that deviate from the HD correlations. In Sec.~\ref{subsec:null_dist}, we will use realistic realizations of an isotropic GWB to derive null distributions that include the effects of fluctuations in cross-correlations. Our results will show that ignoring the impact of cosmic and pulsar variance leads to underestimated $p$-values (i.e., overestimated significance) for anisotropies detection.
    \item \emph{Map reconstruction and detection forecasts}: the likelihood function defined in Eq.~\eqref{eq:likelihood} assumes that the cross-correlations have a Gaussian distribution centered around their ensemble average value. However, we only have access to one realization of the GWB, and for any given realization the cross-correlations will be centered around values that deviate from their ensemble average (as discussed in Sec.~\ref{sec:timing_corr}). In Sec.~\ref{subsec:map_rec}, we investigate how these deviations can affect the recovered sky maps and impact anisotropy detection forecasts. 
\end{itemize}

\begin{figure}[t]
    \centering
    \includegraphics[width=\textwidth]{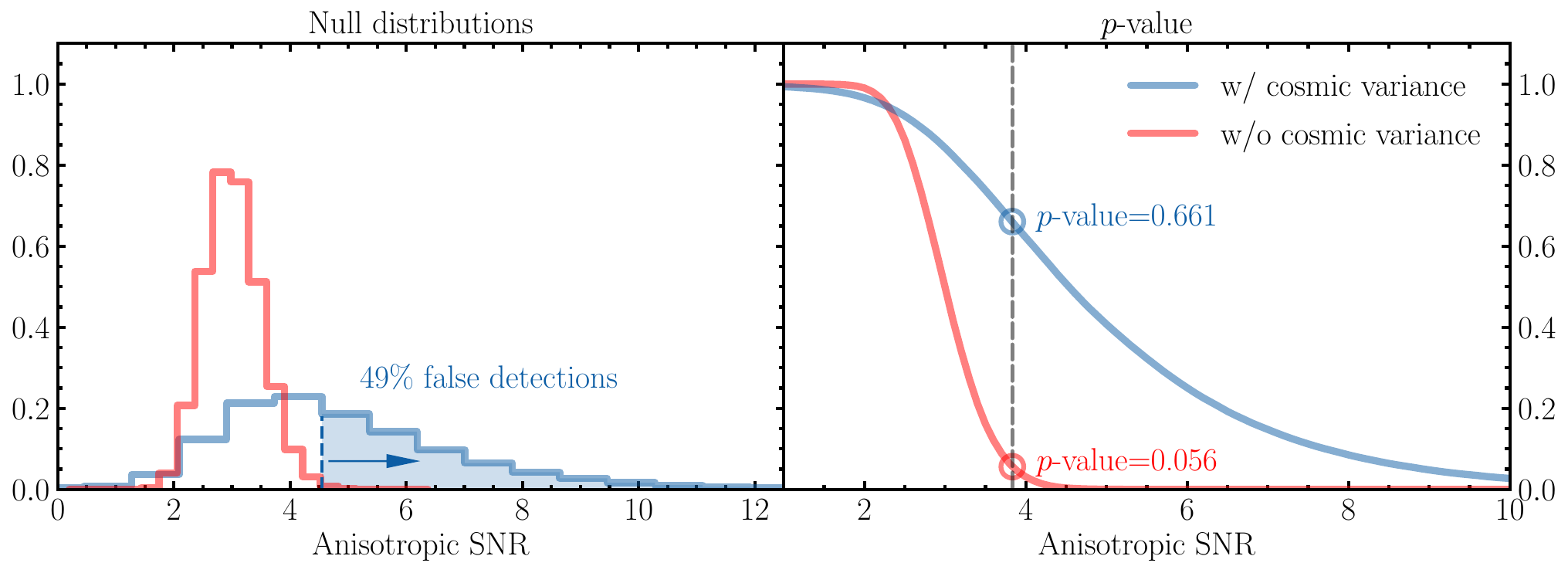}
    \caption{\emph{Left}: null distributions for the anisotropic SNR, defined in Sec.~\ref{subsec:det_stat}, for maps reconstructed using the square-root spherical harmonic basis defined in Sec.~\ref{subsec:det_stat}. The red distribution is derived by assuming that isotropic GWB produces HD correlations, and it closely matches the one derived in Ref.~\cite{NANOGrav:2023tcn}. The blue distribution is derived by including the impact of cross-correlation fluctuations away from the HD correlation induced by GW interference. \emph{Right}: $p$-values as a function of the anisotropic SNR derived with (blue line) and without (red line) including the impact of fluctuations in cross-correlations. The vertical black dashed line corresponds to the anisotropic SNR value measured in the recent NANOGrav anisotropy search~\cite{NANOGrav:2023tcn}.}
    \label{fig:snr_null_astro}
\end{figure}
\subsection{Cosmic variance and detection calibration}\label{subsec:null_dist}
As discussed in the previous section, by itself, the measured value of the detection statistic (whether the anisotropic SNR or individual pixel values) is meaningless. What we are interested in is understanding how likely it is to measure an equal or larger value of the detection statistic under the null hypothesis, i.e., for a realization of an isotropic GWB. We answer this question by comparing the measured detection statistic value with null distributions generated by simulating many realizations of the null hypothesis and computing the corresponding detection statistic. Current anisotropy searches derive such null distributions by assuming that any realization of an isotropic GWB will produce HD correlations.
Therefore, they generate --as realizations of the null hypothesis-- mock correlations by drawing from independent Gaussian distributions with mean given by the HD values and variance given by the measured noise. However, this procedure fails to capture the non-Gaussian and correlated deviations from HD correlations produced by the interference effects discussed in Sec.~\ref{sec:timing_corr}. 
In this section, we study how ignoring these fluctuations impacts the null distributions used to calibrate anisotropy searches. 

\begin{figure}[t!]
\centering
    \includegraphics[width=\textwidth]{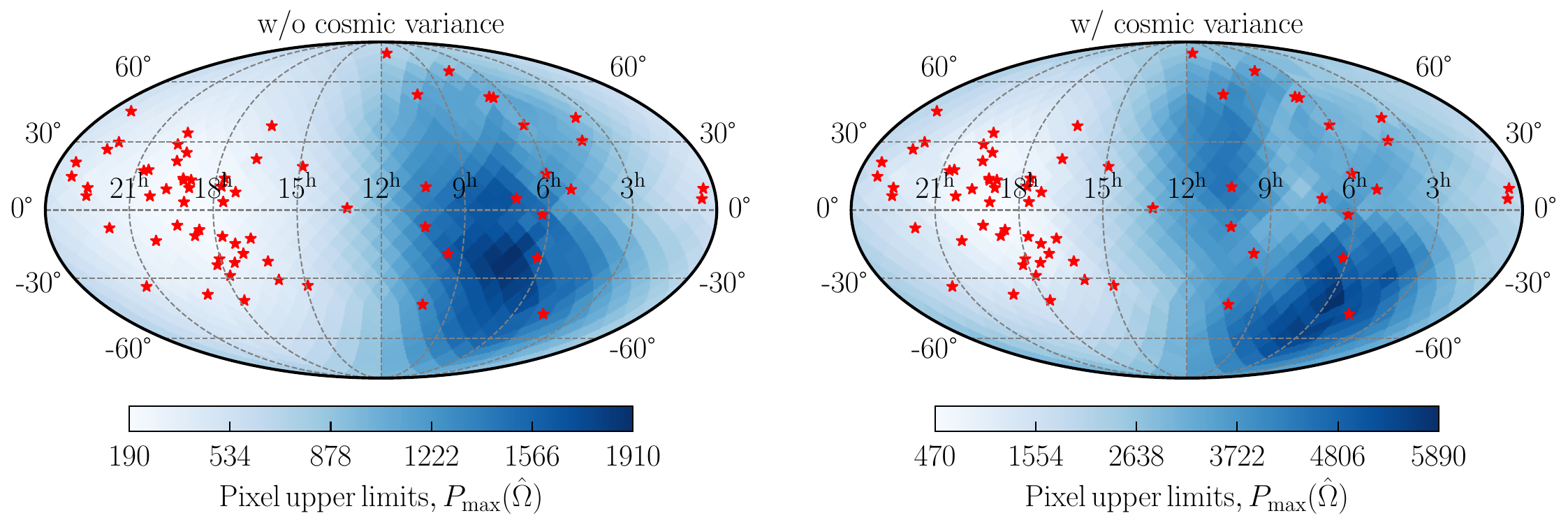}
    \caption{Pixel upper limits derived with (right panel) and without (left panel) the inclusion of cross-correlations variance. The red stars indicate the location of the pulsars contained in the NANOGrav 15-year data set. The numbers outside of the sky map represent the declination angle in degrees, while the numbers on the $x$-axis give the value of the right ascension in hours (1 hour corresponds to $15^\circ$ of sky rotation).}
    \label{fig:sky_upper_astro}
\end{figure}

Specifically, we proceed as follows: we describe an isotropic GWB as a Gaussian ensemble, fully characterized by its power spectrum via Eq.~\eqref{eq:covariance}; we then generate $10^6$ realizations of this ensemble (i.e., $10^6$ realizations of the GW amplitudes and phases, $\{h_{kj}^A,\,\phi_{kj}^A\}$, for each pixel of an \texttt{HEALPix}~\cite{healpix} sky-tessellation with $N_{\rm side}=16$, see App.~\ref{app:gwb_gen} for more details).\footnote{Only $10^5$ of these realizations were used to derive the SNR null distribution shown in Fig.~\ref{fig:snr_null_astro}, while we used all the $10^6$ simulations to derive the pixel upper limits shown in Fig.~\ref{fig:sky_upper_astro}.} For each of these realizations, we calculate the theoretical cross-correlation values using Eq.~\eqref{eq:corr_broad_pix} and then, to simulate realistic reconstruction errors, we add to each of them a random (real) number drawn from a normal distribution with zero mean and standard deviation given by the cross-correlations uncertainties measured in the NANOGrav 15-year data set~\cite{NANOGrav:2023tcn}. We then use these noisy correlations to derive the maximum-likelihood estimator for the GWB sky map following the procedure discussed in Sec.~\ref{sec:ani_search}, and calculate the corresponding detection statistics values (i.e., the anisotropic SNR for maps reconstructed using the square-root spherical harmonic parametrizations, and the individual pixel values for the radiometer maps). We then use these values of the detection statistics to construct their null distributions. For all the results presented in this section, we assume that the power spectrum of the GWB is proportional to $H(f)\propto f^{-7/3}$, the expected frequency dependence for the background produced by a population of SMBHBs in circular orbits, whose evolution is dominated by GW emission.\footnote{The normalization of the GWB spectrum does not play a role in our analysis since we always normalize the cross-correlations by a factor that is proportional to the power spectrum normalization. See the discussion around Eq.~\eqref{eq:norm}.}

The anisotropic SNR null distributions for maps reconstructed using the square-root spherical harmonic parametrizations are reported in the left panel of Fig.~\ref{fig:snr_null_astro}. As we can see from this figure, the null distribution derived using realistic realizations of an isotropic GWB (blue distributions) is wider and peaked at higher SNR values compared to the one derived by assuming that isotropic GWBs produce HD correlations (red distributions). We also find that ignoring the effect of fluctuation in cross-correlations when deriving the null distribution leads to a $\sim49\%$ false-detection rate, where we identify as detection maps with an SNR with a corresponding $p$-value smaller than $3\times10^{-3}$ (which translate to a $\sim3\sigma$ significance). 

In the right panel of Fig.~\ref{fig:snr_null_astro}, we show the $p$-value as a function of the anisotropic SNR. As expected, ignoring the impact of the correlations variance leads to underestimating the $p$-values associated with a given SNR. For example, for the anisotropic SNR measured in the NANOGrav 15-year analysis, SNR $\simeq3.8$, we get a $p$-value $\simeq 0.06$ when ignoring fluctuations in cross-correlations, while $p$-value $\simeq 0.7$ when including them; meaning that realistic realizations of an isotropic GWB are ten times more likely to explain the observed level of anisotropies than previously estimated. 

\begin{figure}[t!]
    \centering
    \includegraphics[width=0.5\textwidth]{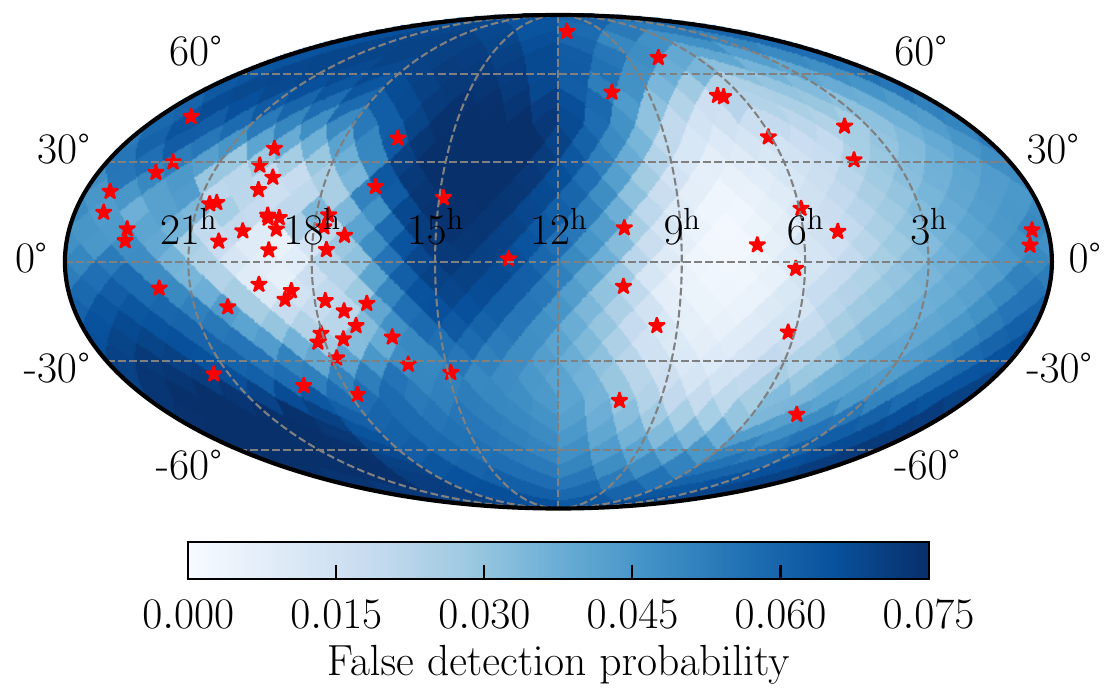}
    \caption{Probability to report a false anisotropy detection in any of the pixels of a \texttt{HEALPix} sky tessellation with $N_{\rm side}=8$ when working in the radiometer basis and deriving the null distributions for the pixel values ignoring the impact of the variance of the cross-correlations.}
    \label{fig:sky_false_det}
 \end{figure}

In Fig.~\ref{fig:sky_upper_astro}, we report the pixel upper limits (obtained by computing the pixel values corresponding to a $p$-value of $3\times10^{-3}/N_{\rm pix}$, which translates to a $\sim3\sigma$ global significance) derived with and without including the impact of cosmic and pulsar variance. From this figure, we can see how using realistic realizations of an isotropic GWB when calibrating the pixel-based detection statistics leads to weaker constraints on the recovered pixel values. We also find that deriving the pixel null distributions without including fluctuations in cross-correlations leads to a false detection rate of $39\%$, slightly lower compared to what we obtained in the square-root spherical harmonic basis. The probability of getting a false detection in a specific pixel is reported in Fig.~\ref{fig:sky_false_det}. 

\begin{figure}[t!]
    \centering
    \includegraphics[width=\textwidth]{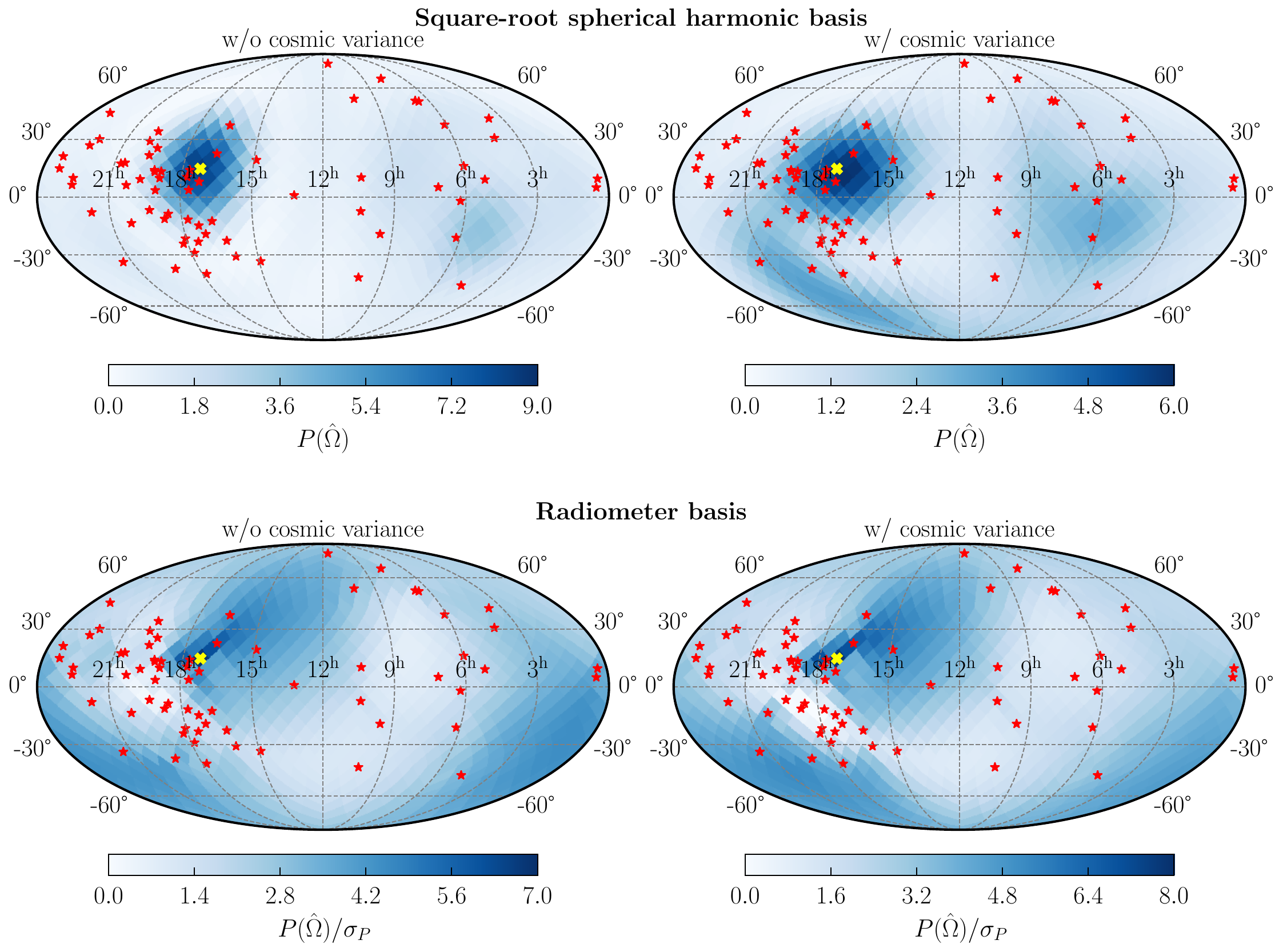}
    \caption{Average of the recovered sky maps for $10^3$ GW skies made of an isotropic GWB plus a hotspot emitting in the first frequency bin. The location of the hotspot (which is kept fixed for all the skies) is indicated with a yellow cross, while the locations of the pulsars are indicated with red stars. The strength of the hotspot is chosen such that $80\%$ of the GW power is in the isotropic GWB and $20\%$ is in the hotspot, which is placed in the first frequency bin at $f_{\rm hot}=1/T$. The pixels of maps recovered using the radiometer basis (lower panels) are normalized using the pixel reconstruction errors, $\sigma_P$. The maps on the left column are reconstructed from the (unrealistic) cross-correlations derived by ignoring any interference effect, while maps on the right column are reconstructed from the (realistic) cross-correlations derived by including these effects.}
    \label{fig:map_rec_1hot}
 \end{figure}

 \begin{figure}
    \centering
    \includegraphics[width=\textwidth]{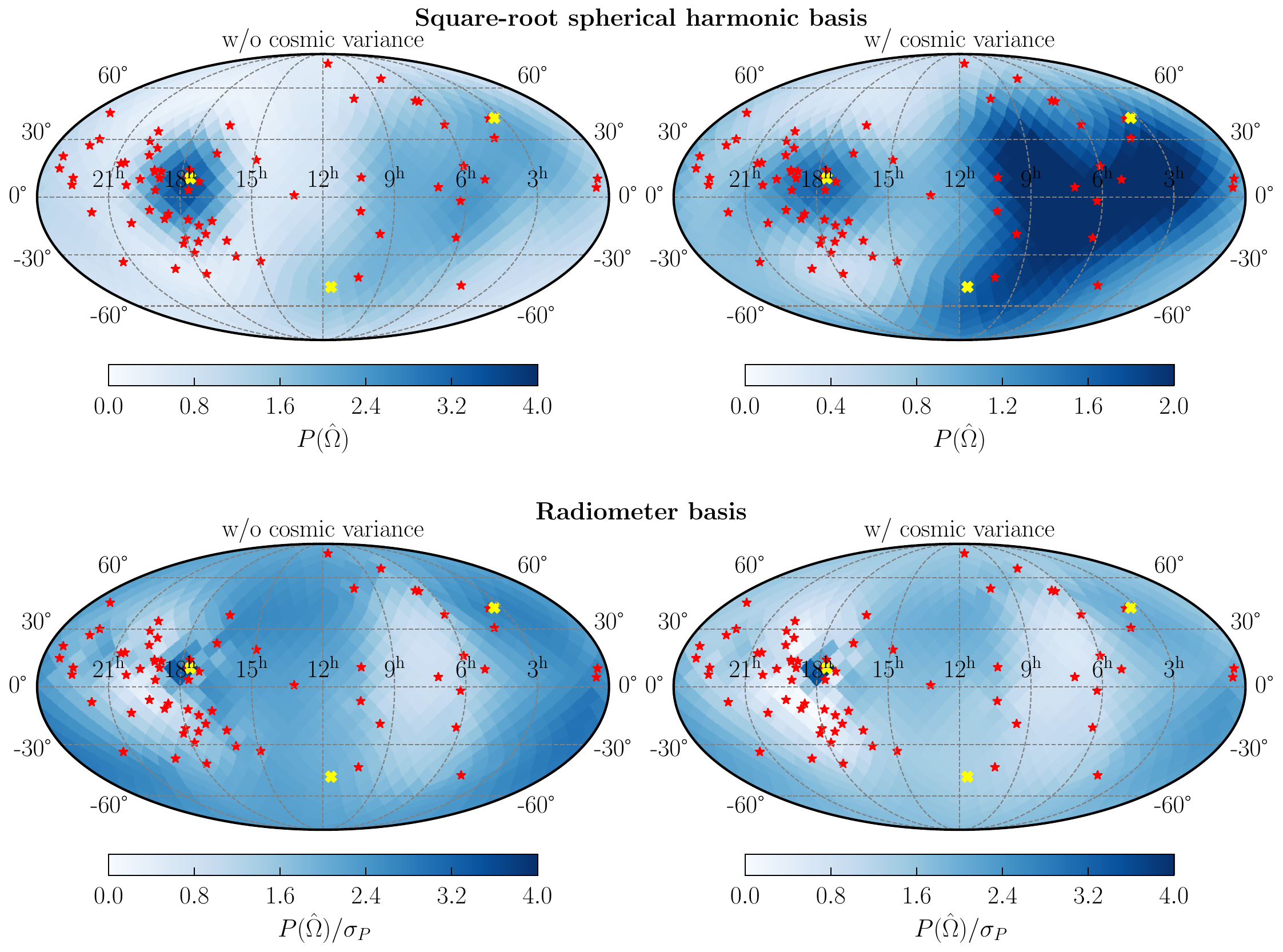}
    \caption{Same as in Fig.~\ref{fig:map_rec_1hot}, but for a sky containing three hotspots, each contributing to $12\%$ of the total GWB power and placed in the first frequency bin. The locations of the three hotspots are indicated by the yellow crosses.}
    \label{fig:map_rec_3hot}
 \end{figure}

 \begin{figure}
    \centering
    \includegraphics[width=\textwidth]{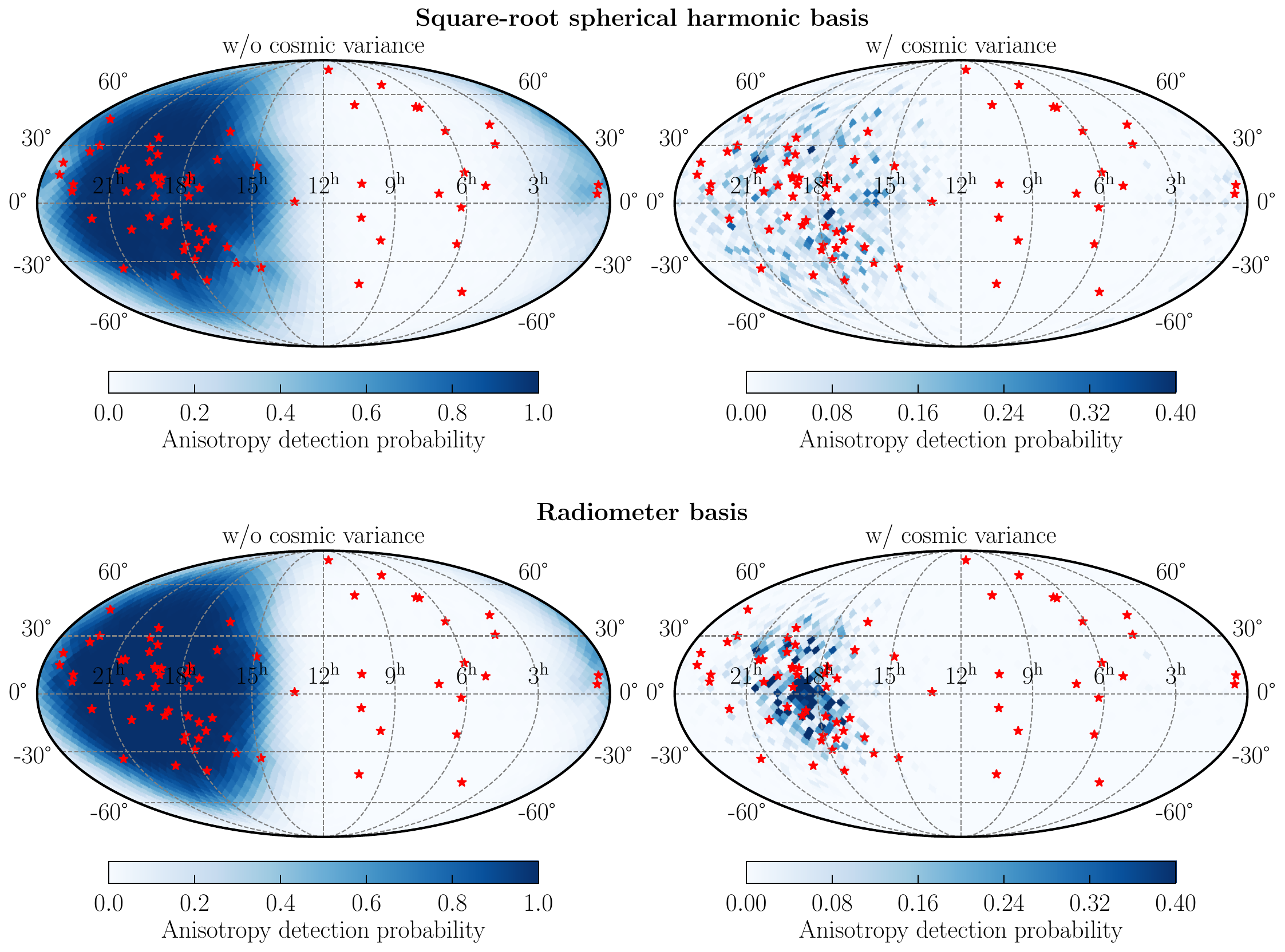}
    \caption{The value of each pixel in these maps reports the anisotropy detection probability for a hotspot, contributing $35\%$ of the total GWB power, placed in that pixel at $f_{\rm hot}=1/T$. The left columns report the detection probabilities derived ignoring the impact of GW interference, while the right columns report the ones derived including these effects. }
    \label{fig:det_pix}
 \end{figure}
 
\subsection{Cosmic variance and map reconstruction}\label{subsec:map_rec}
In this section, we study how cross-correlation variance impacts our ability to localize bright, small-scale GWB hotspots like the ones expected to be produced by a population of SMBHB (see, for example, ~\cite{Taylor:2013esa,Mingarelli:2017fbe}).

To do this, we generate a collection of $10^3$ mock GWB skies, all containing a hotspot on top of an isotropic GWB. We keep the location of the hotspot fixed across realizations and assume that it contributes to $20\%$ of the total GW power and is located in the first frequency bin at $f=1/T$ (i.e., we set the real amplitude of the hotspot pixel $k_{\rm hot}$ such that $(h^{+}_{k_{\rm hot},j_{\rm hot}})^{2}/\sum_{k,j}(h^{+}_{k,j})^{2}=(h^{\times}_{k_{\rm hot},j_{\rm hot}})^{2}/\sum_{k,j}(h^{\times}_{k,j})^{2}=0.2$). We then generate the induced cross-correlations for each of these GWB skies, first ignoring the impact of interference effects (i.e. taking the diagonal component of the $k$ and polarization sum in Eq.~\eqref{eq:corr_broad_pix}), then including these effects. We then add random noise consistent with the one measured in the recent NANOGrav analysis~\cite{NANOGrav:2023tcn} and run these noisy cross-correlations through the reconstruction pipeline discussed in Sec.~\ref{sec:ani_search}.

In the left panel of Fig.~\ref{fig:map_rec_1hot}, we show the average maps reconstructed from the cross-correlations that did not include the interference effects, while the average maps reconstructed from (realistic) correlations by including interference effects are shown in the right panels. We see that, at least for this specific example, the impact of GW interference does not drastically reduce our ability to reconstruct the location of the injected hotspot. This result stems from the fact that, for a single hotspot, interference effects are subleading, as the $k$-sum in Eq.~\eqref{eq:corr_broad_pix} is dominated by the hotspot pixel.
However, interference effects become more relevant for skies containing several hotspots of comparable magnitude. In Fig.~\ref{fig:map_rec_3hot}, we show the average reconstructed map for a sky with three hotspots, each contributing to $12\%$ of the total GWB power and placed in the first frequency bin. In the absence of interference, maps reconstructed using the square-root spherical harmonic parametrization (upper panels) can recover the hotspot in the region of the sky with the highest density of pulsars. However, when we include interference effects, the reconstructed maps miss the location of the hotspots. For the radiometer basis, we find poor reconstruction accuracy with and without the inclusion of interference effects. This result was expected since, by construction, the radiometer basis assumes that only one bright hotspot is dominating the signal. While these results were derived for a specific choice of the hostpost locations, and the impact of interference might be less pronounced for different hotspot locations, they identify scenarios in which ignoring interference effects could hinder our ability to reconstruct the GWB sky. 

All in all, by reducing the map reconstruction accuracy, as just discussed, and increasing the detection thresholds, as discussed in the previous section, the effect of correlations variance is to make it much harder to claim an anisotropy detection using current techniques. To clarify this point, we estimate the anisotropy detection probabilities for a hotspot contributing $35\%$ of the total GWB power and placed in the first frequency bin as a function of its sky position. We do this by repeating the procedure used to derive Fig.~\ref{fig:map_rec_1hot} for each of the pixels of an \texttt{HEALPix} sky tessellation with $N_{\rm side}=16$. However, instead of averaging over the reconstructed maps, for each map, we compute the measured detection statistic (i.e. the anisotropic SNR for maps reconstructed using the square-root parametrization, and individual pixel values for radiometer maps) and by comparing it with the null distributions derived in Sec.~\ref{subsec:null_dist} we determine if that sky would lead to a detection of anisotropies. We then use the fraction of detections in each pixel to estimate the detection probability. The results of this procedure are shown in Fig.~\ref{fig:det_pix}, in the left columns for the case in which we ignore the effect of GW interference, and in the right column for the case in which we include interference effects. It is clear from this figure how interference effects drastically reduce anisotropy detection prospects of current search strategies.

\section{Conclusions and future directions}\label{sec:conclusions}
As the evidence for a gravitational wave background continues to mount, it is crucial to refine the data analysis tools used to characterize its properties. In fact, although the origin of this background remains unknown, improved spatial and spectral characterization provides a path to differentiate between astrophysical and cosmological sources. Specifically, evidence for an anisotropic distribution of the GWB power would strongly suggest an astrophysical origin and disfavor many cosmological sources. Therefore, accurately estimating the statistical significance of any such evidence is essential in the effort to uncover the origin of the GWB.

In this work, we studied how likely it is for a GWB realization drawn from an isotropic ensemble to be erroneously classified as a detection of anisotropies. In current searches, we find that up to $49\%$ of realizations drawn from an isotropic ensemble would lead to a false detection. This false detection happens because, in any given realization of a GWB, the interference between the GW radiation received from different sky directions can significantly impact the pulsars' signals and lead to significant deviations from HD correlations, even for an isotropic GWB. To prevent false detections in future searches, we derive updated null distributions that account for these interference effects for the detection statistic used in current anisotropy searches by the NANOGrav collaboration. These updated null distributions can be used in future searches to prevent false detection rates and better assess the significance of anisotropy detections. 

We also investigate how these interference effects affect our ability to reconstruct the GWB sky map and impact anisotropy detection forecasts. We find that, especially in the case where there are multiple bright sources, interference effects can significantly impact our ability to reconstruct the GW sky accurately. 
For skies with a single hotspot, we find that the higher detection thresholds derived by using the updated null distributions are such that the probability of detecting a bright, small-scale hotspots is severely reduced by interference effects. For skies with multiple hotspots, the degraded sky-reconstruction is expected to limit even further the detection prospects.

Several improvements can be considered to reduce the impact of GW interference on the detection prospects for GWB anisotropies. One first step, that we plan to investigate in a future work, might consist of including the pair-covariance~\cite{Allen:2022ksj} in the off-diagonal components of the covariance matrix appearing in the likelihood function defined in Eq.~\eqref{eq:likelihood}. Another possible improvement could consist of modifying the central value of the likelihood function, Eq.~\eqref{eq:likelihood}, to account for interference effects. This modification would require at least doubling the search parameters, as we would need to reconstruct both the pixel amplitude and the GW phase for each pixel.

Finally, while we studied the detection prospects for single, small-scale GWB hotspots, we plan to revisit the detection prospects for realistic GWB skies produced by SMBHB populations presented in Ref.~\cite{Lemke:2024cdu} to include the interference effects studied in this work. 

\section*{Acknowledgments}
This work was supported by the Deutsche Forschungsgemeinschaft under Germany’s Excellence Strategy - EXC 2121 Quantum Universe - 390833306. We thank Bruce Allen, Kyle A. Gersbach, and Nihan S. Pol for useful discussions and comments on the draft. This work used the Maxwell computational resources operated at Deutsches Elektronen-Synchrotron DESY, Hamburg (Germany).
\appendix

\section{GWB simulation procedure}\label{app:gwb_gen}
In this appendix, we provide additional details on the procedure we used to generate GWB realizations. Generating a realization of a given GWB consists in generating a realization of the complex functions $\tilde h^A(f, \hat{\Omega})$ appearing in Eq.~\eqref{eq:gwb_metric} or, when describing the GWB as a discrete sum of plane-wave, the coefficients $\tilde h_{kj}^A$ defined below Eq.~\eqref{eq:corr_broad_pix}. 

These complex coefficients are treated as random variables, which follow a normal distribution with zero mean and a two-point function given by the discrete version of Eq.~\eqref{eq:covariance}:
\begin{equation}\label{eq:covariance_pix}
    \langle \tilde h^{A*}_{kj}\tilde h^{A'}_{k'j'}\rangle = \delta_{AA'}\delta_{kk'}\delta_{jj'}\frac{H_j}{\Delta
\hat\Omega\,\Delta f}\qquad\qquad\langle \tilde h^{A}_{kj}\tilde h^{A'}_{k'j'}\rangle=\langle \tilde h^{A*}_{kj}\tilde h^{A'*}_{k'j'}\rangle=0\,,
\end{equation}
where $\Delta\hat\Omega$ is the pixel bin-size, and $\Delta f=1/(10T)$ is the frequency bin-size. It is convenient to rewrite the complex amplitudes in terms of a real amplitude, and a complex phase as $\tilde h_{kj}^A\equiv h_{kj}^A e^{i\phi_{kj}^A}$. 
For circularly-symmetric Gaussian random variables, like the one defined by the relations in Eq.~\eqref{eq:covariance_pix}, each of the amplitudes in a given frequency bin is an independent random variable following a Rayleigh distribution, $p(x)$, with a scale parameter given by $\sigma_j^2 = H_j/(2\Delta\hat\Omega\Delta f)$:
\begin{equation}
    p(x) =  \frac{x}{2\pi\sigma_j^2}e^{-x^2/2\sigma_j^2}\,.
\end{equation}
While the phases, $\phi_{kj}^A$, are independent random variables following a uniform distribution between 0 and $2\pi$. 

Therefore, to generate a GWB realization, we generate a set of $\{h_{jk}^A,\phi_{jk}^A\}$ coefficients by drawing $2\times N_f\times N_{\rm pix}$ (one for each pixel, frequency, and GW polarization) independent random variables from a Rayleigh distribution with a frequency-dependent scale factor $\sigma_j$, and an equal number of phases drawing from a uniform distribution between 0 and $2\pi$.
\bibliographystyle{apsrev4-1}
\bibliography{references}

\end{document}